\documentclass[12pt]{article}
\usepackage{epsfig}

\newcommand{\be}{\begin{eqnarray}}
\newcommand{\ee}{\end{eqnarray}}
\newcommand{\sll}{\raise.15ex\hbox{$/$}\kern-.43em\hbox{$l$}}
\newcommand{\slp}{\raise.15ex\hbox{$/$}\kern-.43em\hbox{$p$}}
\newcommand{\slq}{\raise.15ex\hbox{$/$}\kern-.43em\hbox{$q$}}
\newcommand{\slk}{\raise.15ex\hbox{$/$}\kern-.43em\hbox{$k$}}
\newcommand{\slepsilon}{\raise.15ex\hbox{$/$}\kern-.53em\hbox{$\epsilon$}}

\begin{document}

\bibliographystyle{unsrt}
\footskip 1.0cm

\thispagestyle{empty}
\begin{flushright}
INT--PUB 05--05
\end{flushright}
\vspace{0.1in}

\begin{center}{\Large \bf {Production of Forward Rapidity Photons in High 
Energy Heavy Ion Collisions}}\\

\vspace{1in}
{\large  Jamal Jalilian-Marian}\\

\vspace{.2in}
{\it Institute for Nuclear Theory, University of Washington,
Seattle, WA 98195\\ }

\end{center}

\vspace*{25mm}

\begin{abstract}

\noindent We consider production of prompt photons in high energy 
gold-gold and deuteron-gold collisions in the forward rapidity region 
of RHIC ($y \sim 3.8$). In this kinematics, the projectile partons 
typically have large $x_{bj}$ while the target partons are mostly at 
very small $x_{bj}$ so that the primary partonic collisions involve 
valence quarks from the projectile and gluons from the target. We take 
the target nucleus to be a Color Glass Condensate while the projectile 
deuteron or nucleus is 
treated as a dilute system of partons. We show that the photon production
cross section can be written as a convolution of a quark-nucleus scattering
cross section, involving a quark anti-quark dipole, with the Leading Order 
quark-photon fragmentation function. We consider different models of the
quark anti-quark dipole and show that measurement of photons in the 
forward rapidity region at RHIC can distinguish between different 
parameterizations of the dipole cross section as well as help clarify
the role of parton coalescence models in hadron production at RHIC.

\end{abstract}
\newpage

\section{Introduction}

Recent observation of the suppression of produced hadron spectra in high energy
deuteron-gold collisions in the forward rapidity region at RHIC has caused much 
excitement in the field \cite{rhic}. While classical multiple scattering,
appropriate for mid rapidity RHIC, give an enhancement of the $p_t$ spectra 
in nuclear collisions \cite{multi} as compared to spectra in proton-proton 
collisions, the Color Glass Condensate formalism 
predicts a suppression of the $p_t$ spectra in the forward rapidity region 
due to the small $x$ evolution and high density of gluons in the target 
nucleus \cite{forsup}. Recent calculations 
of hadron spectra in deuteron-gold collisions based on Color Glass Condensate 
formalism have been quite successful in describing the RHIC data 
\cite{dima}. More recently, some models based on parton coalescence have 
been used to fit the forward rapidity data \cite{rudi}. While the success of 
these models does not necessarily contradict the Color Glass Condensate 
formalism (since these models use the low transverse momentum data to tune 
their parameters), it is imperative to measure other observable, such as 
dilepton and photons, in order to gain a better understanding of the degree 
to which the Color Glass Condensate is the dominant physics in the forward 
rapidity region and to what degree parton coalescence models are relevant. 
For instance, if photon production in the forward rapidity deuteron-gold 
collisions is suppressed analogously to hadron production, it would 
confirm gluon saturation as the correct physics for particle production
with no need for parton coalescence. Electromagnetic probes such as dileptons 
and photons have the further advantage that they do not interact strongly 
after they are produced and are therefore relatively clean even though 
their production rates are lower \cite{al}. 

In this brief note, we calculate the production cross section for 
photons in the forward rapidity region at RHIC. Starting with the 
production cross section for a quark and a photon \cite{fgjjm}, we 
integrate over the quark transverse momentum and show that this integration 
leads to a divergence which is identified as the collinear divergence
present when a massless quark emits a photon. We show that this 
divergence can be isolated and rewritten as the Leading Order (LO) 
quark-photon fragmentation function \cite{baier1} so that the overall cross 
section can be written as the convolution of a quark-nucleus scattering cross
section \cite{adjjm} and the quark-photon fragmentation function. In order to 
evaluate this cross section, we use the two available parameterizations of the 
dipole-nucleus cross section and show that they lead to quite different
predictions for the ratio of cross sections in deuteron-gold and gold-gold
collisions ($R_{dA}$ and $R_{AA}$). This difference is attributed to the
different shapes of the dipole with respect to its size in the two 
parameterizations. We then discuss the advantages and disadvantages of
the two parameterizations.

In the very forward kinematic region, one probes the large $x$ partons
in the projectile (deuteron or gold) while probing the very small $x$
region in the target nucleus. Therefore, we use the Color Glass Condensate
formalism \cite{glr,nonlin} to describe the target nucleus. On the other hand, 
since the projectile deuteron or nucleus is mainly probed at large $x$, it is
treated as a collection of partons, in this case, quarks and anti-quarks. 
Furthermore, in the case of a deuteron we ignore the nuclear modifications of the 
deuteron wave function since it is a small effect while in the case of 
a gold projectile, we use the EKS98 parameterization of nuclear (anti) 
shadowing \cite{eks}.

\section{The Scattering Cross Section}

The scattering cross section for production of a massless quark with momentum
$q_t$ and a photon with momentum $k_t$ was considered in  \cite{fgjjm} and the 
following 
expression was derived (where $z$ is the fraction of the parent quark carried
away by the photon)
\be
{d\sigma^{q(p)\, A \rightarrow q(q)\,\gamma(k)\, X}
\over dz\, d^2b_t\, d^2k_t} \!\!\!\!&=& \!\!\!\!
{e^2 \over (2\pi)^5} \, z \, [1 + (1-z)^2] 
\, {1\over k_t^2} \, \int d^2 q_t {(q_t + k_t)^2 \over [z q_t - (1-z) k_t]^2}
\, \tilde{N}(x,k_t+q_t,b_t)
\label{eq:cs_gen}
\ee
where $ \tilde{N}(x,k_t,b_t)$ is the dipole cross section in the momentum space
defined as 
\be
\tilde{N}(x,k_t,b_t)= \int d^2 r_t e^{i k_t \cdot r_t} \, N(x,r_t,b_t)
\label{eq:cs_ft}
\ee
and 
\be
N(x,r_t,b_t) = {1\over N_c} \, Tr <1 - V^{\dagger} (x_t) V (y_t)>
\label{eq:cs_def}
\ee
with $r_t\equiv x_t - y_t$ and $b_t\equiv (x_t + y_t)/2$. 
In order to obtain the photon production cross section, we integrate over
the produced quark momentum $q_t$. As is clear from (\ref{eq:cs_gen}), there
is a collinear singularity at $z\, q_t = (1-z)\, k_t$. Shifting 
$z\, q_t \rightarrow z\, q_t + (1-z)\, k_t$ leads to the following
expression for the $q_t$ dependent part of the expression (\ref{eq:cs_gen}),
up to some finite pieces, 
\be
\int_0^{\hat{s}} {d^2 q_t \over q_t^2} \,e^{i q_t \cdot r_t}
\label{eq:div}
\ee
which gives $\pi \log \hat{s}/\Lambda^2$ where $\hat{s}$ is the subpartonic
center of mass energy squared and $\Lambda$ is an infrared cutoff. It is 
common to write this log as a sum of two pieces 
$\log \hat{s}/\Lambda^2 \equiv \log \hat{s}/Q^2 + \log Q^2/\Lambda^2$ where
$Q$ is the factorization scale. The collinear singularity is then absorbed
in the quark-photon fragmentation function $D_{\gamma/q} (z,Q^2)$. The
fragmentation function evolves with the factorization scale $Q$ and obeys 
an evolution equation similar to DGLAP for the parton distribution functions.
At the Leading Log level, this evolution does not change the $Q^2$ dependence
but does change the $z$ dependence. Since we are working in a limited
transverse momentum range, the DGLAP evolution of the fragmentation function
is not important and will be neglected. Furthermore, the fragmentation piece is
the parametrically important one since it is formally of order 
${1\over \alpha_s}$ which cancels a factor of the strong coupling constant 
in the hard cross section so that it is more leading (in $\alpha_s$) than the 
other piece which is sub-leading. Extracting the Leading Order quark-photon 
splitting function  
$P_{\gamma/q} \equiv {e^2 e_q^2 \over 8\pi^2} {1 + (1-z)^2 \over z}$
and identifying 
$D_{\gamma/q} (z,Q^2) \equiv P_{\gamma/q} \, \log Q^2/\Lambda^2$
as the Leading Order quark-photon fragmentation function, the cross section
reduces to
\be
{d\sigma^{q\, A \rightarrow \gamma \, X}
\over dz\, d^2b_t\, d^2k_t} &=& {1\over (2\pi)^2} {1\over z^2}\, 
D_{\gamma/q} (z,k_t)\, \tilde{N}(x,k_t/z,b_t)
\ee
where $z\equiv {k_t\over x_q \sqrt{s}}e^{y_{\gamma}}$ and $x_q$ is the
fraction of the projectile hadron (nucleus) carried by the incoming
quark and $y_{\gamma}$ is the produced photon rapidity. To relate this
to deuteron (nucleus)-nucleus scattering, we convolute the above with
the quark (and anti-quark) distribution functions of the projectile deuteron
or nucleus and sum over the different quark (anti-quark) flavors
\be
{d\sigma^{d(A)\, A \rightarrow \gamma(k_t, y_{\gamma})\, X}\over 
d^2b_t\, d^2k_t\, dy_{\gamma}}\!\!\! &=& \!\!\!
{1\over (2\pi)^2} \sum_f  \int dx_q \,
[q_f(x_q,k_t^2) + \bar{q}_f(x_q,k_t^2)]\, {D_{\gamma/q}(z,k_t^2) \over z}
\tilde{N}(x_g,{k_t\over z},b_t)
\label{eq:cs}
\ee
where $x_g={k_t\over\sqrt{s}} e^{-y_{\gamma}}$ and the lower limit in the $x_q$
integration is $x_q^{min}={k_t\over \sqrt{s}} e^{y_{\gamma}}$. 

We use eq. (\ref{eq:cs}) to calculate the photon production cross 
section in deuteron (gold)-gold collisions. In the case of deuteron-gold 
collisions, we will ignore nuclear modifications of the deuteron since they are 
small in this kinematic region. In the case of projectile gold nucleus, 
we use the parameterization of quark and anti-quark distribution functions 
due to Eskola et al. \cite{eks}. The main effect in the projectile nucleus
wave function is anti-shadowing of quarks and anti-quarks which can be as 
big as $10\%$ in the large $x$ region where the projectile partons are.
Furthermore, we will concentrate on central collisions since this is where
the effects of Color Glass Condensate in nuclei is most prominent so that
our results are meant for photon production in the most central collisions,
for example, $0-5\%$. Generalizing this to more peripheral collisions is
conceptually straightforward but requires a Monte Carlo simulation of centrality
classes which is time consuming and numeric intensive. Since we are mostly
interested in the effects of the Color Glass Condensate on the produced photon
spectra, we will limit ourselves to the most central collisions and leave
the impact parameter dependence of the spectra for a future study.

To proceed further, we need to know the dipole cross section $N(x_g,r_t,b_t)$.
One can in principle solve the JIMWLK equations for $N$ subject to some
initial condition. This has not been accomplished so far since the JIMWLK
equations are highly non-linear, coupled equations. Rather, the large $N_c$
limit of JIMWLK equations (known as the BK equation \cite{bk}) has been 
studied in 
detail and approximate analytical solutions have been found in the high 
energy (large rapidity) limit \cite{bknum}. Alternatively, phenomenological 
models of the dipole cross section, which respect the general properties of the
JIMWLK equations) have been proposed and used to fit the data from HERA
and RHIC. In this work, we use the two known parameterizations of the dipole
cross section proposed in \cite{iim} and \cite{kkt}. We show that the two
models lead to rather different predictions for the dependence of the 
nuclear modification factors $R_{dA}$ and $R_{AA}$ with the photon momentum 
$k_t$ so that, in principle, measurement of the nuclear modification
factor in deuteron (gold)-gold collisions in the forward rapidity region at 
RHIC can shed further light on the detailed dynamics of the Color Glass 
Condensate.

\begin{figure}[htp]
\centering
\setlength{\epsfxsize=8cm}
\centerline{\epsffile{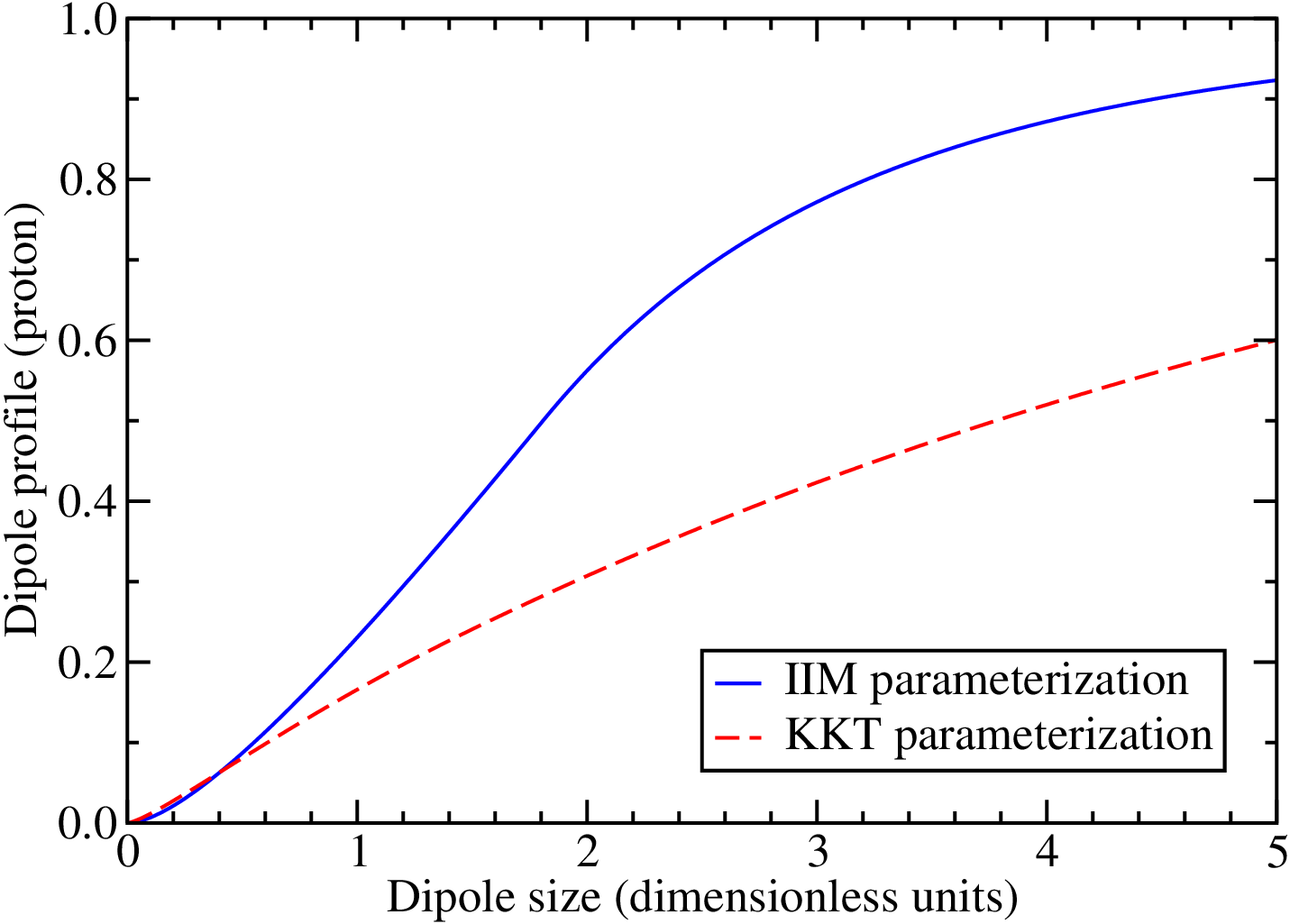}}
\caption{Quark anti-quark dipole profile for a proton target.}
\label{fig:dipole_p}
\end{figure}

\begin{figure}[hbp]
\centering
\setlength{\epsfxsize=8cm}
\centerline{\epsffile{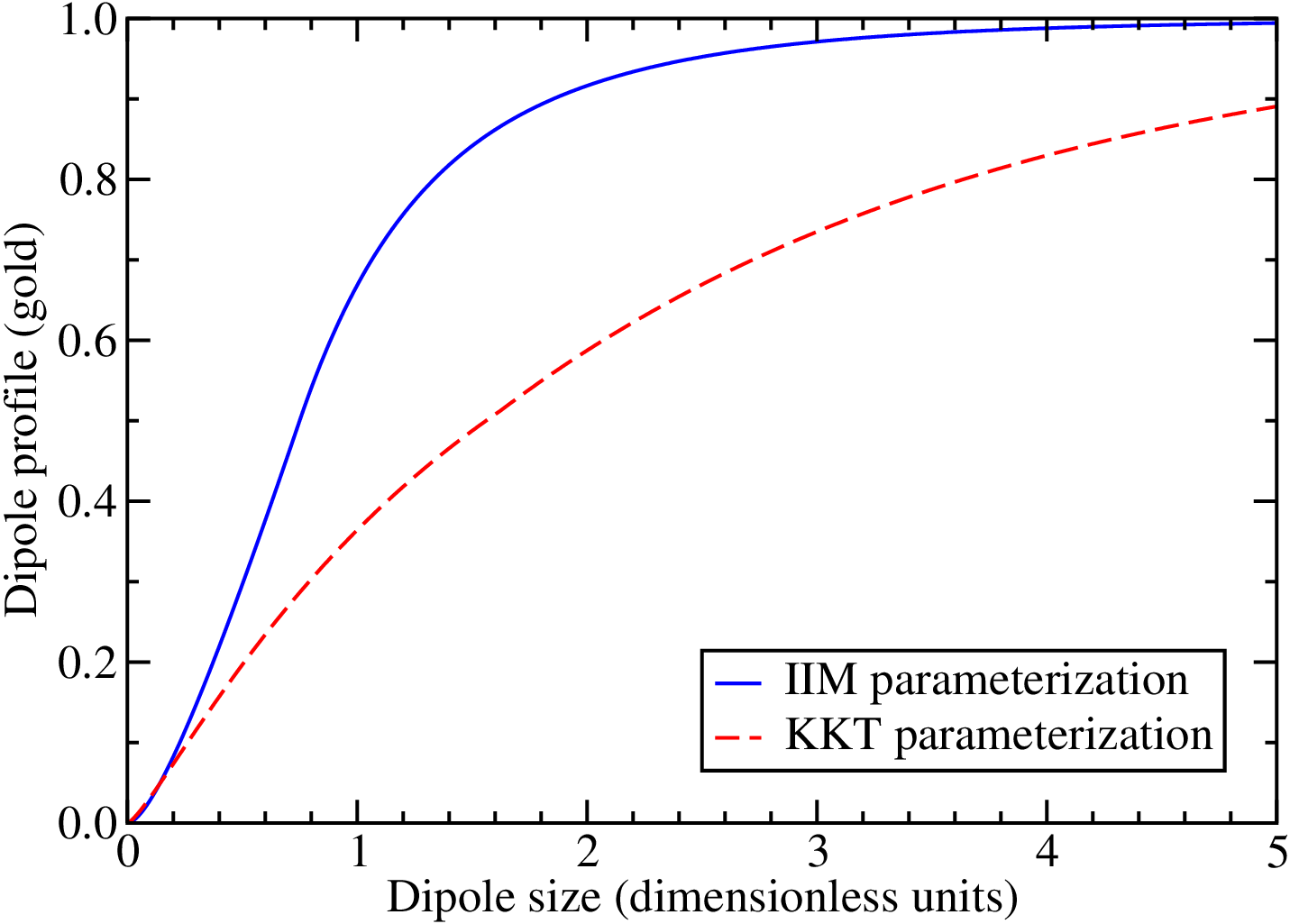}}
\caption{Quark anti-quark dipole profile for a nuclear target.}
\label{fig:dipole_a}
\end{figure}

In Fig. (\ref{fig:dipole_p}), we show the dipole profile for a proton using
the two parameterizations, due to Iancu et al. \cite{iim} (denoted IIM)
and due to Kharzeev et al. \cite{kkt} (denoted KKT) for $x_g=1.6\times 10^{-4}$ 
($k_t=1.5$ GeV at $y_{photon}=3.8$) in terms of the dimensionless parameter
$r_t\,Q_s$. Clearly, the two models lead to quite different results. In Fig. 
(\ref{fig:dipole_a}), we show the dipole profile for a gold target for the 
same values of $x_g$. Again, the two parameterization are quite different.
The profiles shown are for one specific value of $x_g$ on which the profile
depends quite sensitively. In evaluating the production cross section,
one samples different values of $x_g$ where the dipole profiles are different.
Nevertheless, the above figures illustrate the difference in the 
available parameterizations of the dipole profile which leads to 
different predictions for the nuclear modification factors.

\vspace{0.3in}

\vspace{0.3in}

We now use the two parameterizations of the dipole cross section in
coordinate space and Fourier transform them to momentum space and use
the result in (\ref{eq:cs}) to get the photon production cross section
in deuteron-gold and gold-gold collisions. As discussed above, we limit
ourselves to central collisions and forward rapidity. Since the STAR
experiment at RHIC has the capability to measure photons at $y=3.8$, we
evaluate the nuclear modification factor for this rapidity. The 
nuclear modifications factors $R_{dA}$ and $R_{AA}$ are defined as
\be
R_{dA}\equiv {{d\sigma^{dA\rightarrow \gamma\, X} \over dy\, d^2k_t\, d^2b_t}
\over 2 A 
{d\sigma^{pp\rightarrow \gamma\, X} \over dy\, d^2k_t\, d^2b_t}}
\,\,\,\,\,\,\,\,
and 
\,\,\,\,\,\,
R_{AA}\equiv {{d\sigma^{AA\rightarrow \gamma\, X} \over dy\, d^2k_t\, d^2b_t}
\over A 
{d\sigma^{pp\rightarrow \gamma\, X} \over dy\, d^2k_t\, d^2b_t}}.
\ee
The $A$ dependence of our normalization of $R_{AA}$ may look different 
from the commonly used one. However, this is due to our use of EKS
shadowing of projectile quark and anti-quark distributions which are 
normalized to unity in the absence of nuclear effects and because of 
the impact parameter slice $d^2b_t$ which is different for a proton and 
a nucleus.

\begin{figure}[htp]
\centering
\setlength{\epsfxsize=8cm}
\centerline{\epsffile{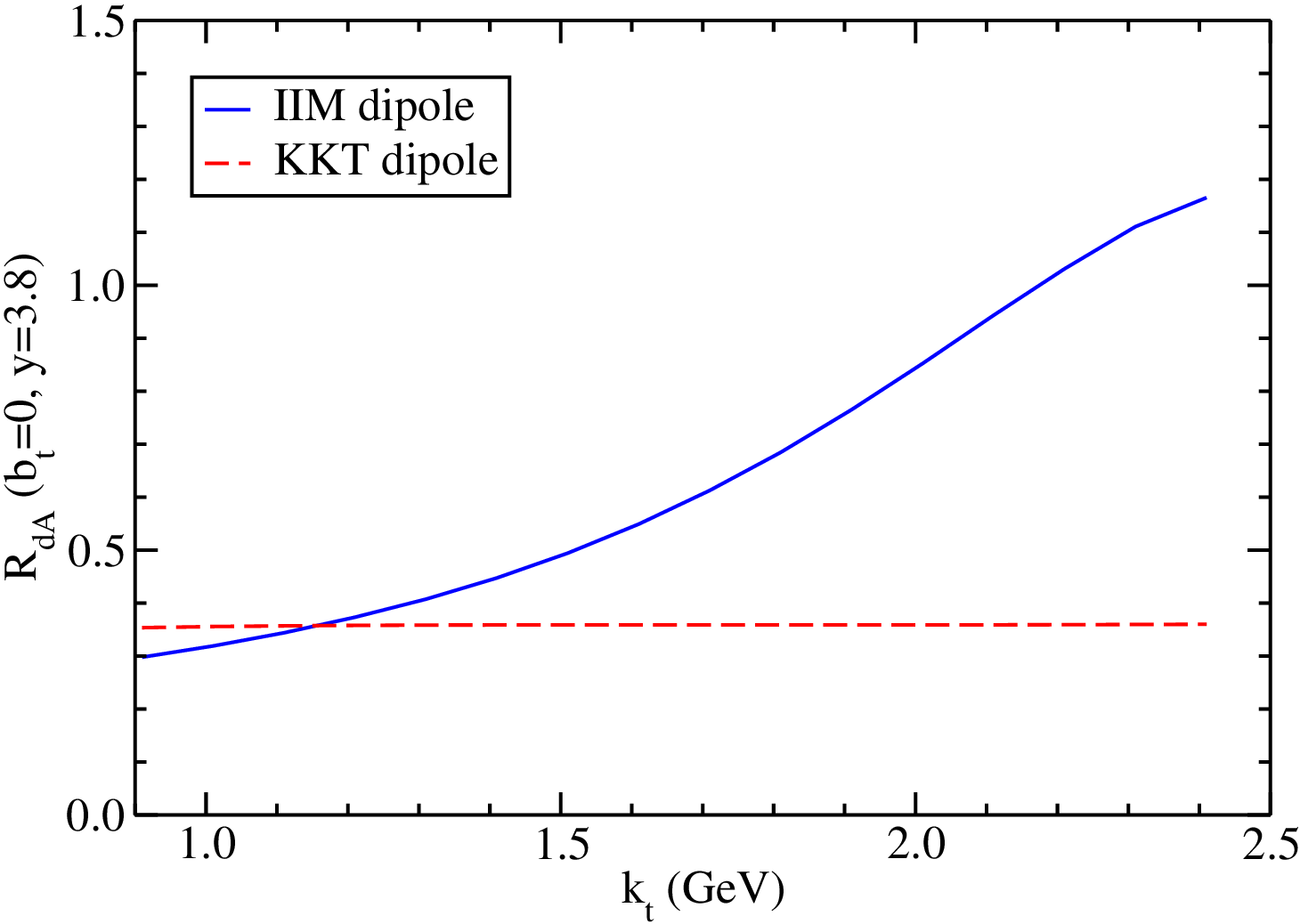}}
\caption{Nuclear modification factor for photon production in 
deuteron-gold collisions.}
\label{fig:R_dA}
\end{figure}

\begin{figure}[htp]
\centering
\setlength{\epsfxsize=8cm}
\centerline{\epsffile{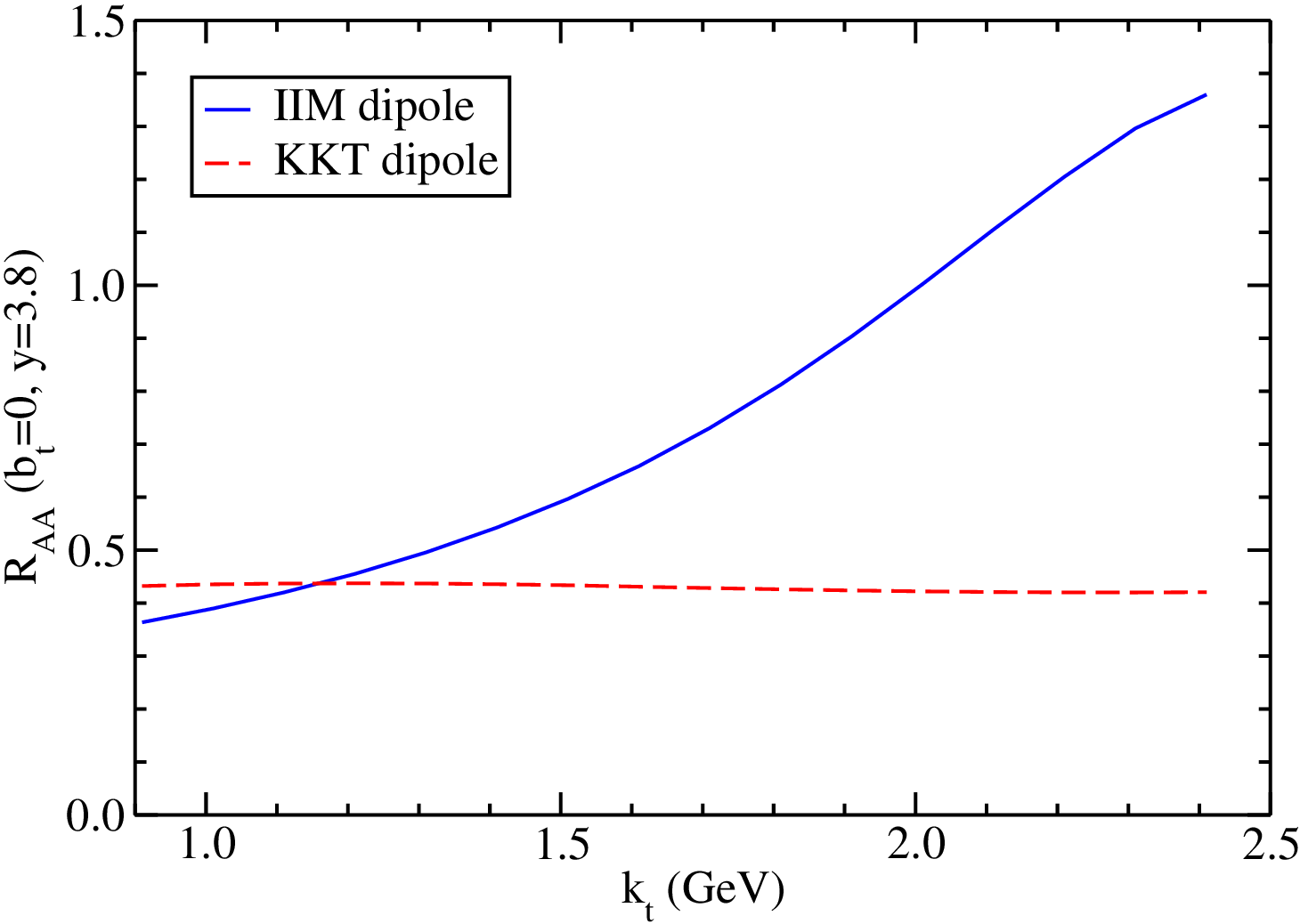}}
\caption{Nuclear modification factor for photon production in 
gold-gold collisions.}
\label{fig:R_AA}
\end{figure}

We show our results for the nuclear modification factors  $R_{dA}$ and $R_{AA}$ 
in Figs. (\ref{fig:R_dA}) and (\ref{fig:R_AA}). The dependence of the 
modification factor on the photon transverse momentum is quite different 
for the two dipole parameterizations due to the different dipole profiles.
The IIM dipole parameterization has a sharper dependence  on its size unlike
the KKT parameterization which grows slower. This is mainly responsible
for the sharp rise of the cross section with $k_t$ in the IIM parameterization.
Another difference between the two parameterization is that the IIM dipole
parameterization does not have the correct high $k_t$ behavior in the double
log region unlike the KKT parameterization which has the right high $k_t$
dependence built in.

\vspace{0.3in}

Furthermore, the quark and anti-quark distributions in the projectile
deuteron and gold are somewhat different due to the anti-shadowing
present in the gold nucleus in the large $x$ region. This causes the
gold nuclear modification factor $R_{AA}$ to be  larger than 
the deuteron nuclear modification factor $R_{dA}$.

\section{Discussion}

Measurement of the nuclear modifications factors for photon production
in the forward rapidity region in deuteron-gold and gold-gold collisions
can help illuminate the presence of the Color Glass Condensate and shed
further light on the detailed dynamics of gluon saturation. The forward
rapidity region is unique in the sense that one probes the smallest
kinematically allowed $x$ in the target and that final state effects
are expected to be negligible, unlike mid rapidity gold-gold collisions 
where final state effects such as the Quark Gluon Plasma are the dominant
effects at RHIC. 

The two
available parameterizations of the dipole cross section, which is the
common element in both hadron and photon production in the forward rapidity
region\footnote{Note that two particle production cross section involves
higher point correlation functions of Wilson loops \cite{bgv,jjmyk}
unlike single particle production which involves only the two point function
of (fundamental or adjoint) Wilson lines.}, have distinctly different 
transverse momentum dependences which can be used to further constrain 
these models. 
Both IIM and KKT parameterizations have advantages as well as disadvantages.
The main advantage of IIM and KKT parameterizations of the dipole
profile, as compared to for instance, the Golec-Biernat-Wusthoff 
parameterization \cite{gbw}, is that these two models have the correct 
anomalous dimension in the low density region which seems to be crucial
for understanding the observed suppression of the forward rapidity data
on hadron production at RHIC. The IIM parameterization 
of the dipole cross section has been made to fit the DIS data on proton
targets in HERA and has not been used for nuclear targets. Our simple
procedure to use it for nuclei by scaling it by a presumed $A^{1/3}$ 
dependence may be too naive since it is not clear that this is valid for 
all dipole sizes. There is a recent study of nuclear DIS data which claims
a $A^{2/3}$ dependence \cite{armesto}. However, the nuclear data are very 
limited in the small $x$ region and the error bars are large. Furthermore,
the DIS structure functions involve a convolution of the dipole cross 
section with the virtual photon wave function squared which weighs
different dipole sizes differently. The KKT parameterization, on the other
hand, is a fit to particle production data at RHIC and has not been checked
against DIS data on proton targets at HERA. Nevertheless, particle
production cross section is a less inclusive quantity than a structure
function measured in DIS and is therefore more constraining even though
the presence of a convolution with the hadron fragmentation function
makes things more non-trivial. Clearly, one needs to check these 
parameterizations of the dipole profile in different observables, such
as photon production considered in this work.

Measuring low $k_t$ photons in the forward rapidity kinematics will be 
challenging. One will need to understand the photon background, due mainly 
to photons coming from neutral pion and $\eta$ meson decays. At the moment,
pions are well measured at RHIC and one can accurately take their 
contributions into account. The same is true for $\eta$ mesons with somewhat less
accuracy. The present data from the last gold-gold run at RHIC should
be precise enough to measure the $R_{AA}$ while another deuteron-gold
may be necessary to get enough statistics in order to extract $R_{dA}$ 
precisely. Another source of photons in the forward rapidity region is
the direct production which become larger than the fragmentation photons
in the high $k_t$ region. In order to estimate the relative contribution
of direct photons, we note that in forward proton-proton collisions at
RHIC, direct photons contribute about $10-15\%$ at $k_t=1$ GeV while
at $k_t = 4$ GeV their contribution is equal to the contribution 
from the fragmentation photons. A precise measurement of the very forward 
rapidity photons at low $k_t$ will go a long way toward establishing the 
Color Glass Condensate as the dominant physics in the forward rapidity 
region at RHIC.

\vspace{0.2in}
\leftline{\bf Acknowledgments} 

We thank L. Blend, A. Dumitru, M. Fontannaz, F. Gelis, Y. Kovchegov, 
S. Kumano and W. Vogelsang for useful discussions. We would also like 
to thank F. Gelis for the use of his Fortran code for the 
dipole cross section and W. Vogelsang for providing us with the invariant cross
section for photon production in proton-proton collisions. This work is 
supported in part by the U.S. Department of Energy under Grant No. 
DE-FG02-00ER41132.
     
\vspace{0.2in}
\leftline{\bf References}

\renewenvironment{thebibliography}[1]
        {\begin{list}{[$\,$\arabic{enumi}$\,$]}  
        {\usecounter{enumi}\setlength{\parsep}{0pt}
         \setlength{\itemsep}{0pt}  \renewcommand{\baselinestretch}{1.2}
         \settowidth
        {\labelwidth}{#1 ~ ~}\sloppy}}{\end{list}}

\end{document}